# Bilayer Photonic Graphene

Mourad Oudich,[1,2,*], Guangxu Su,[3,4,*] Yuanchen Deng,[1] Wladimir Benalcazar,[5] Renwen Huang,[3,4] Nikhil JRK Gerard,[1,6] Minghui Lu,[3,7]  Peng Zhan,[3, 4, †] and Yun Jing[1,†]

[1]Graduate Program in Acoustics, The Pennsylvania State University, University Park, Pennsylvania, 16802, USA

[2]Université de Lorraine, CNRS, Institut Jean Lamour, F-54000 Nancy, France

[3]National Laboratory of Solid State Microstructures, Collaborative Innovation Center of Advanced Microstructures, Nanjing University, Nanjing, 210093, China

[4]School of Physics, Nanjing University, Nanjing, 210093, China

[5]Department of Physics, The Pennsylvania State University, University Park, PA 16801

[6]Department of Mechanical and Aerospace Engineering, North Carolina State University, Raleigh, North Carolina 27695, USA

[7]Department of Materials Science and Engineering, Nanjing University, Nanjing, Jiangsu 210093, China

†yqj5201@psu.edu ; zhanpeng@nju.edu.cn

* M. O. and G. S. contributed equally to this work.

## Abstract

Drawing inspiration from bilayer graphene, this paper introduces its photonic analog comprising two stacked graphene-like photonic crystals, that are coupled in the near-field through spoof surface plasmons. Beyond the twist degree of freedom that can radically alter the band structure of the bilayer photonic graphene, the photonic dispersion can be also tailored via the interlayer coupling which exhibits an exponential dependence on the distance between the two photonic crystals. We theoretically, numerically, and experimentally characterize the band structures of AA- and AB-stacked bilayer photonic graphene, as well as for twisted bilayer photonic graphene with even and odd sublattice exchange symmetries. Furthermore, we numerically predict the existence of magic angles in bilayer photonic graphene, which are associated with ultra-flat bands resulted from interlayer hybridization. Finally, we demonstrate that the bilayer photonic graphene at a particular twist angle satisfying even sublattice exchange symmetry is a high-order photonic topological insulator. The proposed bilayer photonic graphene could constitute a useful platform for identifying new quantum materials and inspiring next-generation photonic devices with new degrees of freedom and emerging functionality.

## I. INTRODUCTION

Over the last decade, the field of van der Waals (vdW) heterostructures has made significant advances towards pushing the frontier of modern materials science. vdW heterostructures are engineered by stacking two-dimensional (2D) atomic layers where the interlayer binding is achieved via weak vdW interaction [1–7]. A well-known family member of vdW heterostructures is bilayer graphene, where two graphene sheets are stacked, introducing an interlayer hopping that results in new electronic and photonic behaviors beyond what is observed in monolayer graphene[8–10]. There are three types of bilayer graphene in view of how they are stacked. They are AA-stacked, AB-stacked, and twisted bilayer graphene (TBG)[10]. In AA-stacked bilayer graphene, each atom of the top layer is placed exactly above the corresponding atom of the bottom graphene. AB-stacked (Bernal) bilayer graphene has received more attention than AA-stacked bilayer graphene due to its greater stability. In such a bilayer graphene, half of the atoms lie directly above the center of a hexagon in the bottom graphene, and half of the atoms lie above an atom. AB-stacked bilayer graphene has been studied for its Landau levels[11], integer quantum Hall effect[12], and quantum valley Hall effect[13,14], among other interesting phenomena[15,16]. As for the TBG, the top graphene layer is rotated with respect to the bottom one by an angle $\theta$, forming a Moiré pattern, which could markedly alter the electronic band of bilayer graphene. The twist degree of freedom possessed by TBG further enriches the physics of bilayer graphene and has enabled TBG to become an important playground for exploring exotic electronic[17-23], optical[9,24-27], and thermal behaviors of vdW heterostructures[16,28,29]. A notable example of TBG is the magic-angle TBG: two stacked graphene sheets misaligned with a discrete set of small twist angles (e.g., 1.08°) that possess flat electronic bands near the Fermi level[17-20,23]. Recent experiments have uncovered that these twist-induced flat bands, remarkably, give rise to unconventional superconductivity[19] as well as correlated (Mott) insulating states[20]. These discoveries have stimulated a host of theoretical and experimental works that attempted to shed light on the underlying physics and potential applications of magic-angle TBG[21,23,26,30-36]. In addition to magic-angle TBG, TBG with large twist angles can be also endowed with extraordinary properties. For instance, it has been found that at certain large twist angles, geometric frustration[32] also gives rise to flat bands, and TBG can serve as a robust candidate material for higher-order topological insulators[34].

On the other hand, there has been a substantial effort in unveiling new physics in classical wave systems through the lens of condensed matter physics. The analogy between these two fields rests on the fact that photonic crystals (PC) or phononic crystals, like crystalline materials, can be also characterized by band structures. This similarity has enabled classical wave researchers to tap into quantum mechanical-like phenomena at a mesoscopic or macroscopic scale. In the context of PCs, recent efforts in this direction include the realization of photonic topological insulators[37-41], Chern insulators[42], valley Hall effects[43,44], Weyl semimetals[45-48], and Landau levels[49]. It is then natural to expect that the spectral properties of bilayer graphene can be also identified in photonics. These properties, however, have been largely unexplored, with only a handful of studies focusing on the theoretical models of AA- and AB-stacked photonic graphene without offering practical designs of PCs for experimental demonstration[50,51]. Notably, a recent study has experimentally investigated a bilayer photonic crystal with mismatched lattice constants[52]. Because the Moiré superlattice therein is not resulted from twist, the physics it entails is fundamentally different from that underpinning TBG. Meanwhile, the twist degree of freedom has been incorporated into the design of optical materials to achieve the localization of light[53] and a topological transition of the isofrequency curve[54-57]. The structures studied in these works, however, are not the exact photonic analogue of TBG. We finally note that the analogue of magic-angle TBG in mechanical wave systems has been recently theoretically investigated[58-60].

The present paper introduces a family of bilayer photonic graphene (BPG) whose band structures and mode behaviors mirror those of AA-stacked, AB-stacked, and TBG, including the magic-angle, sublattice exchange (SE) even, and SE odd bilayer graphene[61]. The underlying working principle of the proposed BPG hinges on the near-field coupling of spoof surface plasmons (SSP). By stacking up two photonic graphene layers with a proper distance in between, the SSPs supported by the two PCs interact with each other to alter the band structure of the monolayer photonic graphene. We theoretically, numerically, and experimentally study the band structures of AA-stacked, AB-stacked, SE even, and SE odd bilayer photonic graphene, all of which show strong resemblance to those of the corresponding bilayer graphene. Additionally, using numerical simulations, we predict the existence of magic angles and topologically nontrivial corner modes in twisted BPG. This exploration of BPG reveals a new route for PC band structure engineering by leveraging the interlayer coupling and the twist degree of freedom. It also establishes a channel between bilayer graphene and PCs, through which the major discoveries in bilayer graphene can be exported across discipline to be presented at a much larger and controllable scale, ultimately facilitating the design of new photonic devices.

## II. RESULTS

### A. AA and AB stacked bilayer photonic graphene

We first consider a monolayer photonic graphene that supports SSP and possesses a Dirac cone at the $K$ point of the Brillouin zone[62] (see **Appendix A**). This PC is made of a hexagonal (graphene-like) lattice of finite length cylindrical metallic pillars decorated on a metallic plate. The distance between any two closest pillars is $a_0$=15 mm (lattice constant $a = a_0\sqrt{3}$); the length and diameter of each pillar are 14.4 and 7.5 mm, respectively. The BPG is subsequently constructed by assembling two layers of photonic graphene in a way that the two PCs face each other while an air-gap of thickness $h$ is kept in between (**Fig. 1(a)**). This configuration allows the SSP in the two layers to couple in the near-field. This interlayer coupling, as will be shown later, is the analog of the interlayer hopping in bilayer graphene. The interlayer coupling strength here can be readily tuned by adjusting the air-gap thickness $h$. This is an important feature and advantage of our photonic system in comparison with bilayer graphene, where the interlayer hopping is far more difficult to tune via pressure or light irradiation[63–65].

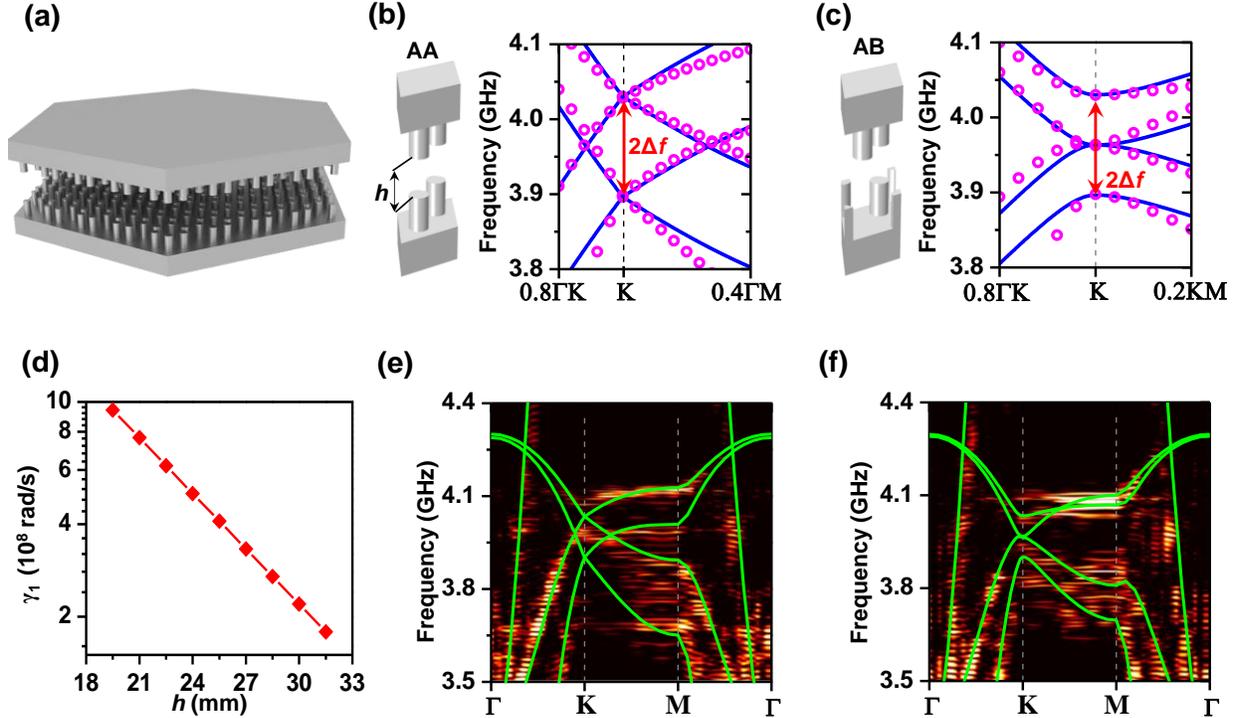

**FIG. 1**. **(a)** Illustration of the BPG composed of two layers of photonic graphene with an air-gap in between. **(b)** and **(c)**, Unit cells for the AA- **(b)** and AB- **(c)** stacked BPG and the associated band structures (solid blue line for theory based on the TBM and pink circles for numerical simulations) in the vicinity of the Dirac frequency for $h$=20.4 mm. **(d)** The interlayer nearest-neighbor coupling strength $\gamma_1$ as an exponential function of the air-gap thickness $h$ (the logarithmic scale is used for the vertical axis). **(e)** and **(f),** Band structures for the AA- **(e)** and AB- **(f)** stacked BPG, obtained numerically (green solid lines) and experimentally (thermal color 2D plot).

We first investigate the AA- and AB-stacked BPG (unit cells are shown in **Figs. 1(b)** and **(c)** by numerically calculating the band structures around the Dirac frequency for the case of $h$=20.4 mm. The band structure of the AA-stacked BPG displays two Dirac cones at the $K$ point separated by a frequency of $2\Delta f$ (**Fig. 1(b)**), while a quadratic dispersion can be seen for the case of the AB-stacked BPG, with two middle bands touching at the Dirac frequency of 3.965 GHz (**Fig. 1(c)**) and the top and bottom bands again being separated by $2\Delta f$. Both dispersion behaviors are similar to the ones observed for the AA- and AB-stacked bilayer graphene[10]. A tight-binding model (TBM) developed in bilayer graphene can be modified to describe the BPG near the Dirac frequency at the $K$ point. The analytical formulation of the Hamiltonian from the TBM is presented in the **Appendix B** and the analytically calculated band structures are compared with the numerical ones (**Figs. 1(b)** and **(c)**). Good agreement is found between the analytical (solid lines) and numerical results (circles). Furthermore, we use the band structure of the AB-stacked BPG to investigate the interlayer nearest-neighbor coupling strength: $\gamma_1 = 2\pi\Delta f$ ($\gamma_1$ is commonly used in bilayer graphene to denote the interlayer nearest-neighbor hopping, following the notion of the Slonczewski-WeissMcClure model[10]). **Figure 1(d)** presents $\gamma_1$ as a function of the air-gap thickness $h$, where an exponential decay of the coupling strength is observed as $h$ increases. This is expected because SSP exhibits evanescent decay from the interface. We constructed a prototype to measure the band structures of the AA- and AB-stacked BPG (see

Appendix C for the experimental setup). The measured band structures are shown in **Figs. 1(e)** and **(f)** where the numerically modeled band structures are overlaid on top of the measurements. The agreement, overall, is reasonable. The discrepancy between the numerical and experimental results can be partially attributed to the fact that some of the modes were not excited in the experiment. The finite size of the sample could also be a contributing factor.

## B. Magic-angle bilayer photonic graphene

For TBG, it has been shown that the bands of the two Dirac cones associated with each graphene would hybridize and form flat bands at a set of discrete angles, denoted the magic angles[17]. Consequently, the Dirac velocity in theory vanishes at these magic angles [17,20]. In order to identify the magic angles in BPG, the band structure of the Moiré unit cell around the Dirac frequency was computed and its bandwidth was evaluated: a vanishing bandwidth at the $\Gamma$ point indicates the emergence of flat bands[23]. Note that twisting one graphene layer with respect to the other generally creates the Moiré pattern with quasiperiodicity. This pattern, however, becomes perfectly periodic for specific values of twist angles known as commensurate angles, which can be determined by [22],

$$\theta_{m,n} = arg\left(\frac{me^{-i\pi/6} + ne^{i\pi/6}}{ne^{-i\pi/6} + me^{i\pi/6}}\right), \qquad (1)$$

where $(m, n)$ are positive integers. Consequently, the band structure can only be rigorously calculated for Moiré unit cells at these commensurate angles. To this end, we first choose a commensurate angle of 3.89° (angles smaller than this value yield larger unit cells which make the calculation of the band structure computationally prohibitive) and vary the interlayer coupling strength by changing the air-gap thickness $h$. This strategy allows us to identify the appropriate interlayer coupling strength at which flat bands arise and consequently renders 3.89° the magic angle. **Figure 2(a)** shows the Moiré lattice and associated unit cell. The band structure is calculated by COMSOL and the results are shown in **Fig. 2(b)** for three values of air-gap thickness ($h$=18.6, 20.4 and 22.2 mm). In particular, the four bands closest to the Dirac point frequency (in red) become flat when $h$=20.4 mm. On the other hand, the four bands are separated at the $\Gamma$ point for higher or lower values of $h$. **Figure 2(c)** displays the eigenstates at the top band near the $\Gamma$ point, indicated by black circles (1), (2) and (3) in the band structure of the three samples shown in **Fig. 2(b)**, respectively. Strong electrical field localization can be observed at the AA stacked regions located at the four corners of the unit cell. To better illustrate how these flat bands evolve with the interlayer coupling strength, the bandwidths of the four bands (the frequency difference between the top and bottom bands) at the $\Gamma$ point near the Dirac frequency are extracted for various air-gap thicknesses. In addition to the twist angle of 3.89°, another commensurate angle 4.409° has also been considered, and their corresponding results are shown in **Fig. 2(d)** (circles, blue for $\theta$=4.409° and red for 3.89°). To validate these results, we have also estimated the bandwidth using an analytical formulation derived originally for TBG[23],

$$BW = \frac{2w}{\alpha}\left(1 - 2\alpha + \frac{\alpha^2}{3} + \cdots\right), \qquad (2)$$

where $\alpha$ is related to the twist angle by

$$\alpha = w/2vk_D \sin(\theta/2). \qquad (3)$$

$k_D$ is the magnitude of wave vector at the $K$ point for the monolayer graphene, $v$ is the Dirac velocity which was estimated to be 2.15×10$^7$ m/s by the TBM for the monolayer photonic graphene (see **Appendix B**), and $w = A\gamma_1$ where $A$ is a fitting parameter. The theoretical prediction of the bandwidth using Eq. 2 is presented by solid lines in **Fig. 2(d)**. A fitting parameter of $A = 1/3$ allows good agreement between the prediction and numerical results (circular dots). This value of $A$ matches with the one previously reported in TBG[66]. This plot suggests that, for the twisted BPG with the air-gap thicknesses of 20.4 and 19.5 mm, the corresponding magic angles are 3.89° and 4.409°, respectively. It is noted that both these magic angles are currently inaccessible in TBG due to the lack of means to achieve the needed interlayer hopping strength. This highlights the strength of our twisted BPG at accessing large magic angles, which are associated with small unit cells and therefore also small device sizes. On the other hand, to attain a magic angle close to the one in magic-angle bilayer graphene (~1.08°), the interlayer coupling strength must be reduced (by increasing $h$) and it is estimated that the air-gap thickness $h$ would be roughly 29.7 mm in the twisted BPG.

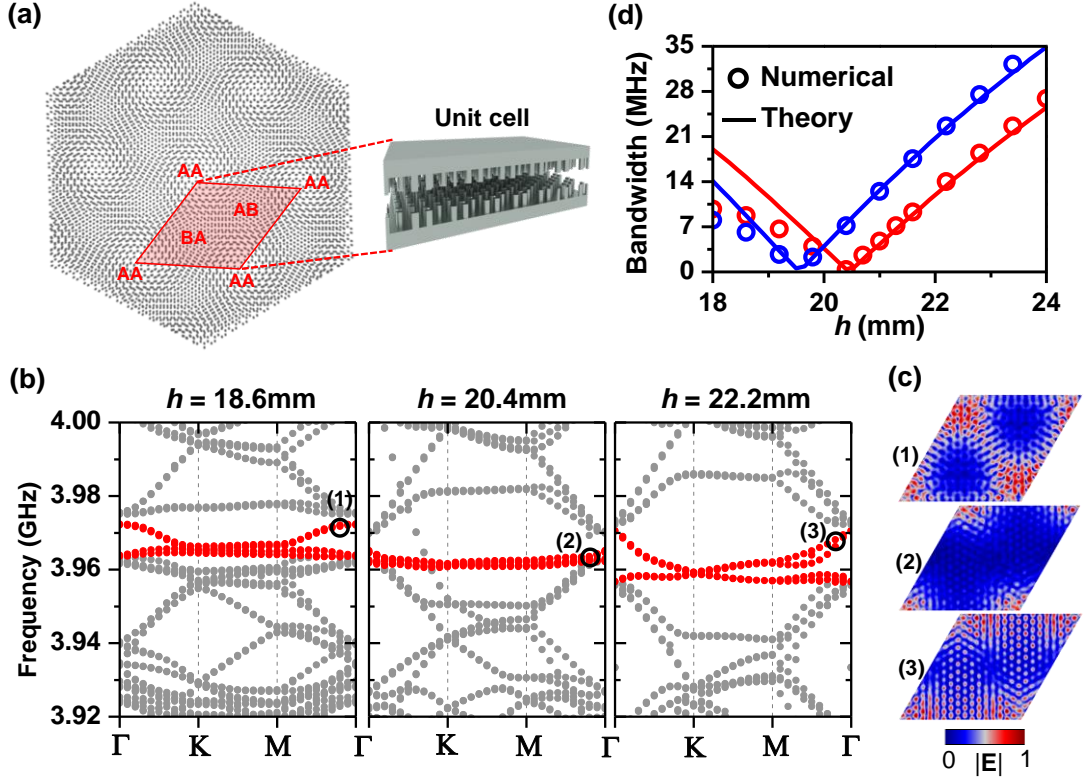

**FIG. 2. (a)** A twisted BPG (Moiré lattice) and its unit cell. The AA, AB, and BA regions are marked out. **(b)** Band structures around the Dirac frequency for the case of air-gap thicknesses $h$ of 18.6, 20.4 and 22.2 mm. The twist angle is θ =3.89°. **(c)** Electric field eigenstates for the modes indicated by (1), (2) and (3) near the $\Gamma$ point in **(b)**. **(d)** Bandwidth of the red bands in **(b)** at the $\Gamma$ point as a function $h$ for the twist angles of 3.89° (red) and 4.409° (blue). Circles represent numerical results and lines represent theoretical results.

## C. SE-even and SE-odd bilayer photonic graphene and topological corner modes

Twisted BPG with large commensurate angles near 30° have also been investigated in this study. This section focuses on the SE symmetries with two commensurate angles $30^o \pm 8.213^o$, where the rotation center is a pillar. The two unit cells associated with the twisted BPG are schematically presented in **Figs. 3(a)** and **(b)** for the SE even and odd symmetries, respectively. For the sake of clarity, we also indicate the positions of the pillars of the upper and bottom layers in red and blue circles, respectively, in a planar representation of the unit cell. For $38.213^o$ ($m = 1, n = 4$), the lattice shown in **Fig. 3(a)** has a six-fold symmetry and the unit cell has coincident pillars in the two layers at the vertices of the hexagonal unit cell (**Fig. 3(a)**). The two layers also share the same hexagon center at the center of the unit cell. In this case, the SE symmetry is even[61]. The case of $21.787^o$ ($m = 1, n = 2$) is of odd SE where the lattice has a three-fold symmetry with coincidence of the pillars between the two layers only at three vertices of the unit cell (**Fig. 3(b)**), while pillar of one layer (red circle at center) is aligned with the hexagon center of the other layer inside the unit cell. Using the unit cell for each of these two SE symmetries with $h$ being 20.4 mm, the band structure around the Dirac frequency can be numerically calculated, and the results are presented in **Figs. 3(c)** and **(d)** (circles). One can see that the photonic dispersion of the SE even and odd cases are similar to the electronic dispersion of the TBG with the same SE symmetries (e.g., see Fig. 3 in Ref [61] and Fig. 2a in Ref [67]). The SE even BPG yields a narrow band gap spanning from 3.95 to 3.954 GHz while the SE odd BPG has a quadratic dispersion with the middle two bands touching at the Dirac frequency, similar to the case of AB-stacked BPG (**Fig. 1(c)**) but with a different frequency range. We also used the TBM (see **Appendix B**) to analytically compute the band structures, which are compared with the numerical ones (solid lines vs. circles in **Figs. 3(c)** and **(d)**). Good agreement can be observed between the two models. A close inspection on the electrical wave field at the edges of the bandgap (P1 and P2 in **Fig. 3(c)**) further reveals that the gap arises from the pseudospin[32,61] characterized by rotational out-of-plane electrical field within the unit cell (**Fig. 3(e)**). Furthermore, for the case of SE odd (**Fig. 3(d)**), the electrical fields at the first and fourth band at the *K* point (C1 and C2 in **Fig. 3(d)**) show chiral modes with rotational out-of-plane electrical field with opposite chirality within the unit cell (**Fig. 3(f)**). We performed experimental characterization of the band structure for SE even and odd BPG near the Dirac frequency and the measurement results are displayed in **Figs. 3(c)** and **(d)** (thermal color map). Though the resolution of the measured band structure is relatively poor in the momentum space, due to the lack of unit cells in the SE even and odd cases, one can still observe the bandgap for the SE even BPG and the gaplessness in the SE odd case.

We have also investigated the topology of the bands below the bandgap shown in **Fig. 3(c)** (see **Fig. 6** in **Appendix A** for the full band structure), for the SE even BPG, and conclude that this bilayer PC can harbor topologically nontrivial corner modes. We diagnose this nontrivial topology by determining the $C_{2z}$ symmetry representations of the 14 modes below the bandgap at the $\Gamma$ and $M$ points. For all the 14 bands below the band gap, the eigenvalues were calculated using the operator

$$\Xi = p(k)C_{2z}p(k), \qquad (4)$$

where $p(k)$ is a projector operator into the 14 bands below the gap at the high-symmetry momenta $k = \Gamma, M$ defined as

$$p(k) = \sum_{i=1}^{14}|\varphi_i(k)><\phi_i(k)|, \qquad (5)$$

where $|\varphi_i(k)>$ is the eigenstate associated with each band $i$ at the high-symmetry momentum $k$, and $<\phi_i|$ is the transconjugant of $|\varphi_i>$. As a result of the projection, the operator $\Xi$ has 14 non-vanishing eigenvalues $\lambda_p = +1$ or $-1$.

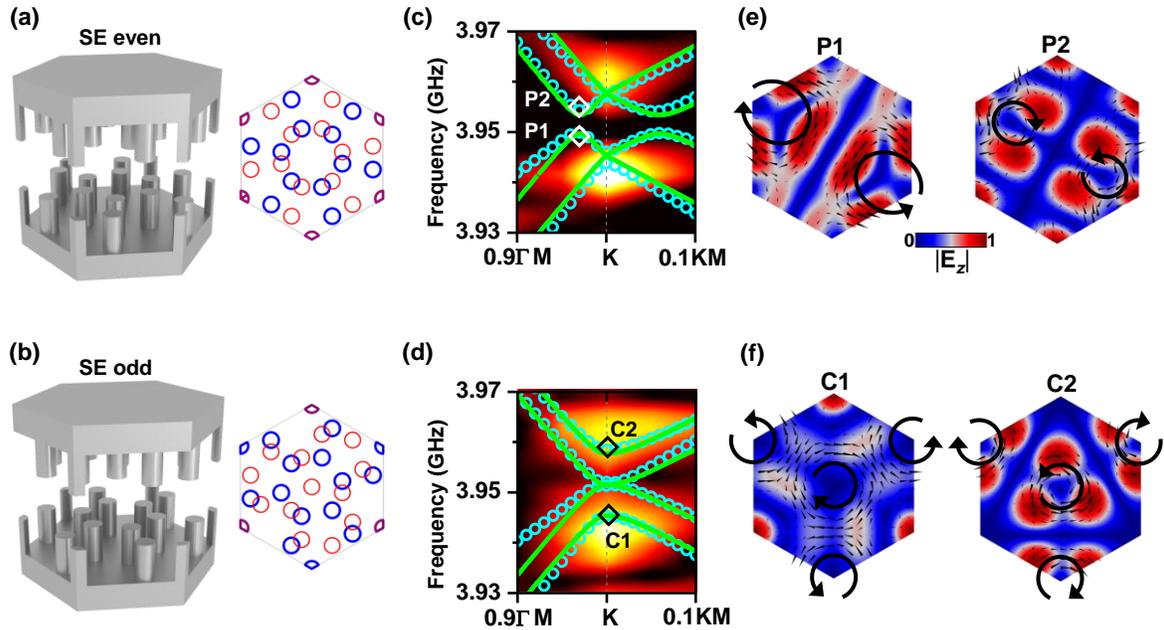

**FIG. 3. (a)** and **(b)**, The unit cells of the twisted BPG for the case of SE even (**a**) and odd (**b**) symmetries. The right panels show the planar representation of the positions of the pillars (red and blue circles for the top and bottom layers, respectively). **(c)** and **(d)**, Band structure around the Dirac frequency for SE even and odd, respectively. Solid green line: theory based on the TBM; cyan circles: simulation; thermal color map: measurement. **(e)** and **(f),** The out-of-plane component of the electrical field $E_z$ for the points marked as P1, P2, C1, and C2 in **(c)** and **(d)**, respectively. The black arrows indicate the energy flow direction characterized by the Poynting vector.

**Table 1** shows the nonvanishing eigenvalues of the operator Ξ at the *Γ* and *M* points. The mismatch in the number of representations between the *Γ* and *M* points is consistent with the topology of an obstructed atomic limit having 3 Wannier centers at the maximal Wyckoff positions c, c′, and c″ of the unit cell[68,69] (**Fig. 4(a)**), while the rest of Wannier centers are fixed by symmetry to the Wyckoff position 1a , i.e., at the center of the unit cell. The overall configuration of Wannier centers results in the existence of nontrivial corner states[69] as a consequence of the existence of corner-localized charges of ½ at the 120° corners (**Fig. 4(b)**).

**Table 1.** Eigenphase counting at the high symmetry points *Γ* and *M*

|  | *Γ* | *M* |
|---|---|---|
| Number of $\lambda_p = -1$ | 7 | 8 |
| Number of $\lambda_p = +1$ | 7 | 6 |

To further assess the non-trivial topology of the bands, the existence of topological corner modes is numerically demonstrated by studying a finite size sample consisting of 12×12 units (**Fig. 4(d)**) with open boundaries. **Figure 4(c)** shows the mode frequencies at the vicinity of the band gap (blue shaded region) where a mode can be found inside the band gap (indicated by the blue diamond dot). The electrical field distribution associated with this mode is plotted in **Fig. 4(e)** which shows energy localization at both corners with 120° angle. Furthermore, we have investigated the robustness of the corner mode by introducing a defect into the sample at both corners (**Fig. 4(g)**). The eigenfrequency calculation (**Fig. 4(f)**) shows that the corner modes subsist in the presence of this defect (**Fig. 4(h)**).

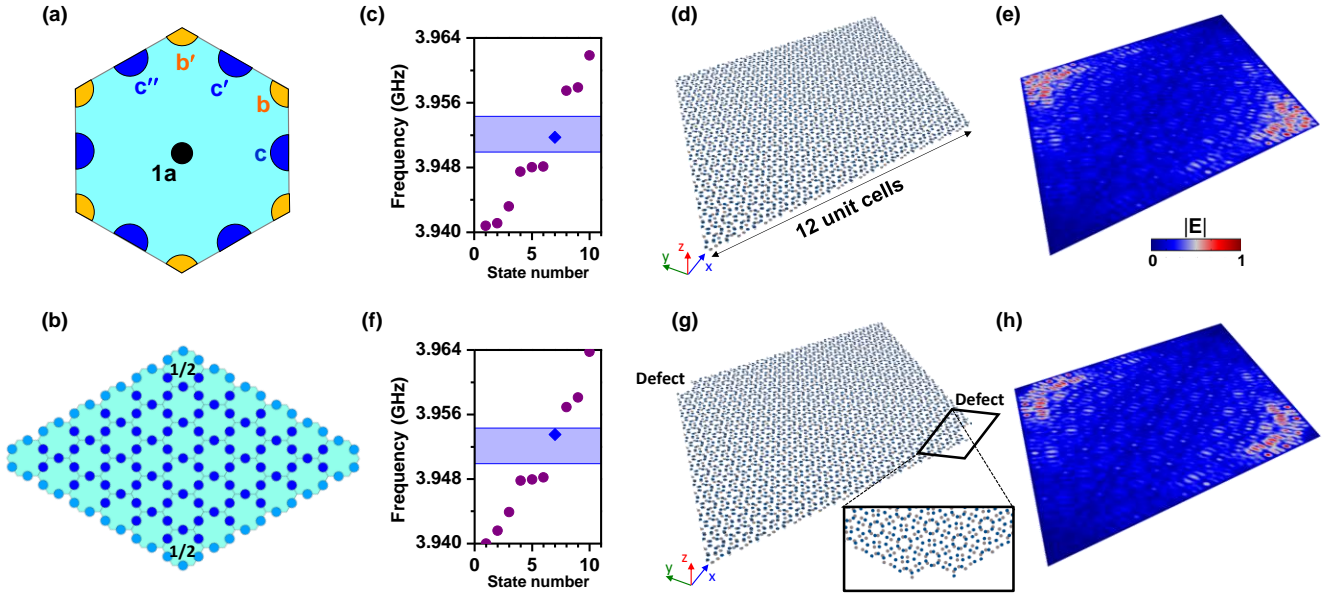

**FIG. 4.** **(a)** and **(b),** Maximal Wyckoff positions in a unit cell with $C_6$ symmetry **(a)** and in a sample of 6 × 6 unit cells **(b)**. The dark blue circles represent bulk Wannier centers while the light blue circles represent boundary positions. **(c)** and **(e),** Eigenfrequencies **(c)** of the sample with 12 × 12 unit cells **(d)** near the Dirac frequency. The electrical field distribution **(e)** shows the corner mode located inside the band gap (blue diamond in **c**). **(f)** and **(h),** Eigenfrequencies **(f)** of the sample with a defect **(g)** constructed by removing 16 pillars at each of the 120° corners (inset). The electrical field distribution **(h)** shows that the corner mode (blue diamond in **(f)**) can be preserved. The shaded regions in **(c)** and **(f)** correspond to the band gap.

### III. CONCLUSIONS

By stacking up two photonic graphene sheets that support interacting SSP, a photonic analogue of bilayer graphene has been proposed, constructed, and characterized. We have investigated theoretically, numerically, and experimentally the dispersion of the BPG with AA and AB stacking, and of the twisted BPG with SE even and odd symmetries. The band structures are shown to mirror those observed in the corresponding bilayer graphene. Furthermore, we have presented, for the first time, a photonic analog of the magic angle originally discovered in TBG, where bands near the Dirac frequency become flat with a vanishing effective velocity. These flat bands give rise to energy localization that can potentially enhance wave-matter interaction and sensing. In addition to the twist degree of freedom, the proposed twisted BPG enables the tuning of the magic angle by tailoring the interlayer coupling, which is associated with the airgap thickness between the two PCs. This feature of BPG unlocks large magic angles that are so far inaccessible in TBG. Finally, we show that the SE even BPG possesses a bandgap which can be exploited to create topologically protected corner modes. Overall, the proposed BPG is expected to serve as a new playground for investigating new physics arisen from the bilayer graphene community as well as for inspiring novel photonic devices and identifying new quantum materials.

## ACKNOWLEDGMENTS

Y. J, Y. D and O. M would like to thank the startup funds from the Pennsylvania State University. P. Z and M. L are financially supported by National Key R&D Program of China (No. 2018YFA0306200, 2017YFA0303702) and National Natural Science Foundation of China (No. 11674166 and 11834007). W. A. B. thanks the support of the Eberly Postdoctoral Fellowship at the Pennsylvania State University.

## APPENDIX A: BAND STRUCTURES OF MONOLAYER PHOTONIC GRAPHENE AND SE-EVEN BILAYER PHOTONIC GRAPHENE

### 1. Monolayer photonic graphene

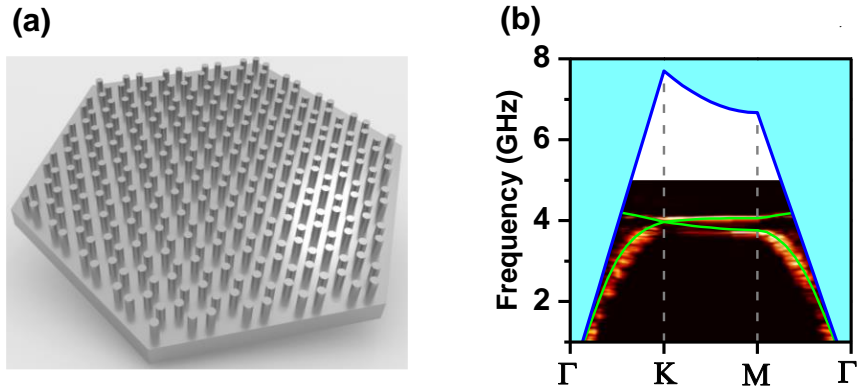

**FIG. 5.** Schematic of a monolayer photonic graphene (**a**) that supports SSP and its associated band structure (**b**) with a Dirac cone at the *K* point, computed numerically (green solid lines) and experimentally (thermal color 2D plot).

We consider a monolayer photonic graphene (PC) that possesses a Dirac cone at the *K* point of the Brillouin zone[62]. The PC is made of a hexagonal (graphene-like) lattice of finite length cylindrical metallic pillars decorated on a metallic plate (**Fig. 5(a)**). It is reasonable to consider the boundaries of this PC to be perfect electric conductors. This PC is known to support SSP in the air domain surrounding its boundaries[72]. **Figure 5(b)** presents the band structure of this monolayer PC, in which a Dirac cone is situated at 3.395 GHz. The numerically modeled band structure (green curve) is overlaid on top of the measurement result with good agreement. In this example, the distance between any two closest pillars is $a_0$=15 mm (lattice constant $a = a_0\sqrt{3}$); the length and diameter of each pillar are 14.4 and 7.5 mm, respectively. The calculation was performed using the commercial software COMSOL Multiphysics 5.4. Because SSP have a slow group velocity, they are located below the light cone (line delimiting the blue shaded region in **Fig. 5(b)**) and the electromagnetic wave field is confined within the air surrounding the pillars at the Dirac cone, due to the evanescent nature of SSP[72].

## 2. Band structure of the SE even bilayer photonic graphene

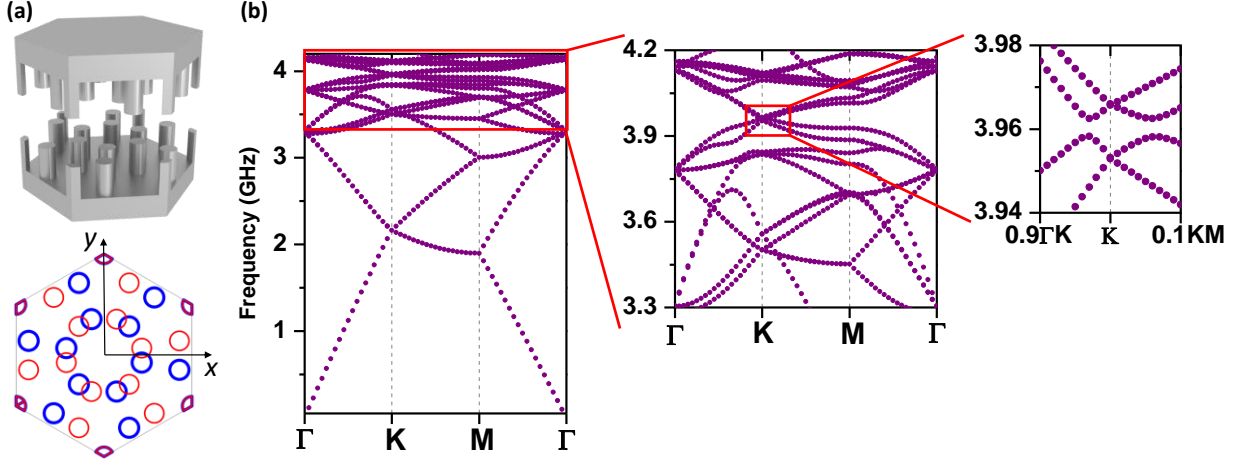

**FIG. 6. (a)** Unit cell of the SE even bilayer (twist angle of θ = 38.213°). The red and blue circles indicate the pillars of the top and bottom layers, respectively. **(b)**, Band structure of the SE even bilayer photonic graphene with successive enlargement near the Dirac frequency.

## APPENDIX B: COMPUTATIONAL METHODS

### 1. Numerical simulation

The numerical characterizations of the BPC were all performed using the commercial software Comsol Multiphysics v5.4. The band structures were calculated by considering the unit cell describing the periodic lattice with Floquet periodic conditions at the appropriate boundaries. The boundaries of the metallic plate and pillars in contact with air were considered as perfect electric conductors, while the refractive index of air is $n_{air}$=1. For the corner mode characterization, a sample of 12×12 unit cells of the SE even lattice where the scattering boundary condition was adopted for the outside open boundaries.

### 2. Tight binding model

The band structures of the monolayer photonic graphene, and the AA- and AB-stacked BPG can be analytically described by the tight-binding model in the vicinity of the *K* point. For the monolayer photonic graphene, the Hamiltonian can be written as,

$$H_{Monolayer} = \begin{pmatrix} \omega_0 & -\gamma_0 f(\mathbf{k}) \\ -\gamma_0 f^*(\mathbf{k}) & \omega_0 \end{pmatrix}, \tag{6}$$

where

$$f(\mathbf{k}) = e^{ik_y a/\sqrt{3}} + 2e^{-ik_y a/2\sqrt{3}} \cos(k_x a/2). \tag{7}$$

Here, $a$ is the lattice constant, $\omega_0$ is the Dirac frequency, and $\mathbf{k} = (k_x, k_y)$. The in-layer coupling parameter $\gamma_0 = 2.87 \times 10^9$ rad/s was estimated by fitting the numerical band structure (from COMSOL) around the *K* point for the monolayer photonic graphene, where the fitting was based on minimizing the root-mean-square deviation.

For the AA-stacked BPG, the following Hamiltonian (slightly modified from the one used in bilayer graphene[70]) can be used

$$H_{AA} = \begin{pmatrix} \omega_0 & -\gamma_0 f(\mathbf{k}) & \gamma_1 & -\gamma_3 f^*(\mathbf{k}) \\ -\gamma_0 f^*(\mathbf{k}) & \omega_0 & \gamma_4 f(\mathbf{k}) & \gamma_1 \\ \gamma_1 & \gamma_4 f^*(\mathbf{k}) & \omega_0 & -\gamma_0 f(\mathbf{k}) \\ -\gamma_3 f(\mathbf{k}) & \gamma_1 & -\gamma_0 f^*(\mathbf{k}) & \omega_0 \end{pmatrix}. \quad (8)$$

For AB-stacked BPG, the Hamiltonian reads

$$H_{AB} = \begin{pmatrix} \omega_0 & -\gamma_0 f(\mathbf{k}) & -\gamma_4 f(\mathbf{k}) & -\gamma_3 f^*(\mathbf{k}) \\ -\gamma_0 f^*(\mathbf{k}) & \omega_0 & \gamma_1 & \gamma_4 f(\mathbf{k}) \\ \gamma_4 f^*(\mathbf{k}) & \gamma_1 & \omega_0 & -\gamma_0 f(\mathbf{k}) \\ -\gamma_3 f(\mathbf{k}) & \gamma_4 f^*(\mathbf{k}) & -\gamma_0 f^*(\mathbf{k}) & \omega_0 \end{pmatrix}, \quad (9)$$

where the interlayer nearest-neighbor coupling $\gamma_1$ follows $\gamma_1 = 2\pi\Delta f$ (**Fig. 1(b)** and **(c)**). The effects of the interlayer hopping terms $\gamma_3$ and $\gamma_4$ were neglected[58,71] in this study.

The Hamiltonians for the SE odd and even BPG, respectively, are[61]

$$H_{\text{SE odd}} = \begin{pmatrix} \omega_0 & -\gamma_0^{SE} f(\mathbf{k}) & \gamma_1 & 0 \\ -\gamma_0^{SE} f^*(\mathbf{k}) & \omega_0 & 0 & 0 \\ \gamma_1 & 0 & \omega_0 & -\gamma_0^{SE} f(\mathbf{k}) \\ 0 & 0 & -\gamma_0^{SE} f^*(\mathbf{k}) & \omega_0 \end{pmatrix}, \quad (10)$$

and

$$H_{\text{SE even}} = \begin{pmatrix} \omega_0 & -\gamma_0^{SE} f(\mathbf{k}) & \gamma_1 e^{i\varphi} & 0 \\ -\gamma_0^{SE} f^*(\mathbf{k}) & \omega_0 & 0 & \gamma_1 e^{-i\varphi} \\ \gamma_1 e^{-i\varphi} & 0 & \omega_0 & -\gamma_0^{SE} f(\mathbf{k}) \\ 0 & \gamma_1 e^{i\varphi} & -\gamma_0^{SE} f^*(\mathbf{k}) & \omega_0 \end{pmatrix}, \quad (12)$$

where $\gamma_0^{SE} = 3.44 \times 10^8$ rad/s was estimated by fitting the numerical band structure (from COMSOL). $\varphi$ is directly related to the width of the SE even band gap (**Fig. 3(c)**), which was found to be $\varphi = 35\pi/180$ through fitting.

## APPENDIX C: FABRICATION AND EXPERIMENTAL CHARACTERIZATION

### 1. Construction of the bilayer photonic graphene

The bilayer photonic graphene was constructed via lathe (**Fig. 7**) using aluminum. The aluminum pillars were 14.4 mm tall and had a diameter of 7.5 mm. The distance between closest pillars was 15 mm. The pillars were arranged over the aluminum plate to form a hexagon shape with an overall diagonal of $30a_0 = 450$ mm, which corresponded to the distance between the two farthest pillars in the sample. The only difference between the top and bottom photonic graphene was the edge of the aluminum plates. The bottom photonic graphene had a hexagonal edge while the top photonic graphene was slightly larger and had a rounded edge. For the monolayer measurement, only the bottom photonic graphene was used during the measurement.

### 2. Experimental setup

The self-built microwave near-field scanning system mainly consisted of two parts: a vector network analyzer (Agilent E5063A) and a 3D electronically controlled displacement stage. A microwave antenna (as an excitation source) was mounted on the bottom aluminum plate through a drilled hole, which was close to the center of the bottom photonic graphene. The motor-controlled probe antenna was inserted laterally into the airgap between the two aluminum plates for scanning. The step size of the scanning was 3 mm. After measuring the amplitude and phase of the electric field in the frequency regime of our interest (from 1.0 GHz to 5.0 GHz), the band structure can be obtained by the spatial Fourier transform, which was processed by MATLAB. During the experiment, the bottom photonic graphene was placed on a platform and the top photonic graphene was held up by three supporting columns at the edge (**Fig. 7**). The top photonic graphene was adjusted to the proper orientation manually to form different BPG configurations. Extreme caution was given to align the center of the two photonic graphene. The twist angle was confirmed by a protractor and the error was ensured to be less than 0.5°.

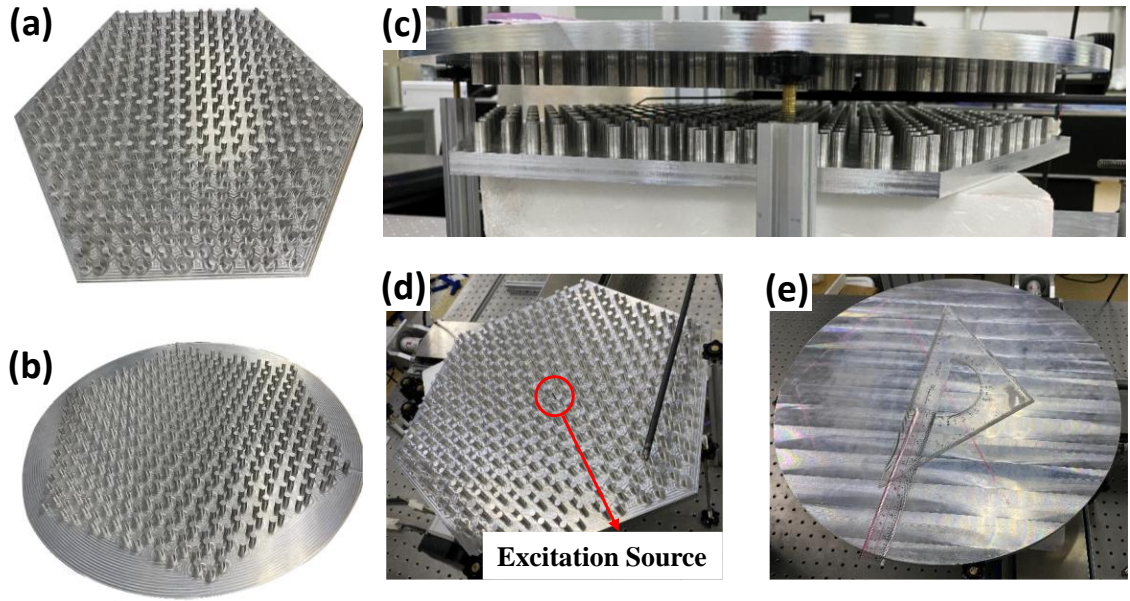

**Fig. 7.** (**a**) and (**b**), Photos of the top and bottom photonic graphene, each made of a hexagonal lattice of pillars decorating an aluminum plate. (**c**), The stacking configuration of the bilayer photonic graphene with an airgap in between the two layers to introduce and control the interlayer coupling. (**d**), bottom photonic graphene where a dipolar source is placed near the center of a hexagon. (**e**), top view of the top photonic graphene where the twist angle is measured by a protractor.